\documentclass[runningheads]{llncs}
\usepackage{amsmath}
\usepackage{hyperref}
\usepackage{graphicx}
\usepackage{svg}
\usepackage{adjustbox}
\usepackage{booktabs}
\usepackage{multirow}
\usepackage{enumitem}
\usepackage{xspace}

\DeclareMathOperator*{\argmax}{arg\,max}
\newcommand{\algoname}[1]{{\textsc{#1}}\xspace}
\newcommand{\ourmodel}{\algoname{DESIRE-ME}}

\begin{document}
\title{DESIRE-ME: Domain-Enhanced Supervised Information REtrieval using Mixture-of-Experts}
\titlerunning{DESIRE-ME}
%
\author{Pranav Kasela\inst{1,2}\orcidID{0000-0003-0972-2424} \and
Gabriella Pasi\inst{1}\orcidID{0000-0002-6080-8170} \and
Raffaele Perego\inst{2}\orcidID{0000-0001-7189-4724} \and
Nicola Tonellotto\inst{2,3}\orcidID{0000-0002-7427-1001}}
\authorrunning{Kasela et al.}
\institute{University of Milano-Biocca, Milan, Italy\\
\email{\{pranav.kasela,gabriella.pasi\}@unimib.it}\\ \and
ISTI-CNR, Pisa, Italy\\
\email{\{raffaele.perego\}@isti.cnr.it} \and
University of Pisa, Pisa, Italy\\
\email{\{nicola.tonellotto\}@unipi.it}}
\maketitle
\begin{abstract}
Open-domain question answering requires retrieval systems able to cope with the diverse and varied nature of questions, providing accurate answers across a broad spectrum of query types and topics. 
To deal with such topic heterogeneity through a unique model, 
we propose \ourmodel, a  neural information retrieval model that leverages the Mixture-of-Experts framework to combine multiple specialized neural models. We rely on Wikipedia data to train an effective neural gating mechanism that classifies the incoming query and that weighs the predictions of the different domain-specific experts correspondingly. This allows \ourmodel to specialize adaptively in multiple domains.
Through extensive experiments on  publicly available datasets, we show that our proposal can effectively generalize domain-enhanced neural models.  \ourmodel excels in handling open-domain questions adaptively, boosting  by up to $12\%$ in NDCG@10 and $22\%$ in P@1, the underlying state-of-the-art dense retrieval model. 

\end{abstract}
\section{Introduction}

The Information Retrieval (IR) research landscape has been fundamentally reshaped by the rapid adoption and emergence of neural models, generating a new paradigm known as Neural Information Retrieval (NIR). Within this transformation, one prominent application of neural models within IR systems is achieved through dense retrieval techniques that have shown promising results in situations where understanding the semantic context of queries and documents is crucial for accurate retrieval.
In  contrast to their traditional counterparts, which heavily rely on lexical similarities captured by scoring functions such as TF-IDF or BM25, dense retrieval techniques naturally capture query and document semantics and can be easily adapted to handle multi-modal data and cross-lingual retrieval~\cite{mitra2019nir}.
However,  their training requires large labeled datasets, and the resulting models  are typically highly specialized to the task they are trained on and do not generalize well to a new task or domain without additional fine-tuning. 

Numerous efforts have been directed towards creating a single neural model that can generalize across many domains, but achieving this goal has proven challenging~\cite{thakur2021beir}.
In attaining this objective, we must also consider that the queries in many IR tasks are often brief and concise, sometimes lacking sufficient information for comprehensive semantic matching. Moreover, users typically do not explicitly specify the domain of their query, so, if necessary, the system must deduce it in a latent manner. 

A sub-field of neural IR is open-domain Q\&A, where the questions are posed in natural language and the answer is retrieved from an extensive collection of documents.
In this work, to address the above issues, we propose \ourmodel, a model for open-domain Q\&A that can specialize in multiple domains without changing the underlying pre-trained language model. 
This specialization is achieved by adaptively focusing the retrieval on the current query domain by  leveraging the Mixture-of-Experts (MoE) framework~\cite{jacobs1991moe}. The MoE framework provides a machine learning architecture  combining multiple specialized models, called ``specializers'' or ``experts'',  to collectively solve a task, such as Q\&A. Each specializer within the framework is designed to excel in a specific topical subdomain or under certain conditions, and the MoE model dynamically selects and combines these specializers to make predictions tailored to the input data. 
A gating mechanism determines which specializer(s) to use for a given input. This gating mechanism is a trained neural network that takes the input query and assigns an importance weight to each expert. The weights indicate the relevance of each specializer for the current input and determine their contribution to the final prediction.
The \ourmodel approach applied to a complex and faceted task such as open-domain  Q\&A  permits learning a robust and adaptive MoE model that  handles the heterogeneity of questions better than state-of-the-art monolithic dense retrievers.
To summarize, our research contributions are:
\begin{itemize}
    \item A modular MoE framework for open-domain Q\&A integrated into a  dense retrieval system that  significantly boosts the performance of the underlying model by exploiting domain specialization;
    \item A supervised gating method able to understand the query topic and correspondingly weighting the domain contextualization computed by the various MoE specializers; 
    \item A novel experimental framework exploiting the folksonomy of Wikipedia to derive automatically the domains of documents and queries used to train the supervised gating mechanisms;
\end{itemize}
We evaluate our proposal against state-of-the-art baselines with reproducible experiments on three different datasets \footnote{The code is available at this link: \url{https://github.com/pkasela/DESIRE-ME}.}. 
The results of the experiments show that \ourmodel consistently improves the performance of the underlying dense retriever with an increase  of up to 12\% in NDCG@10 and $22\%$ in P@1, outlining the potential of the proposed model for the open-domain Q\&A task. 
Furthermore, we utilize a fourth dataset having similar characteristics to investigate the generalization capabilities of \ourmodel in a zero-shot scenario. Even in this case, we observe a significant performance boost over the underlying dense retriever.

The paper is organized as follows. Section \ref{sec:related} discusses the relevant related work. Section \ref{sec:method} formally introduces the \ourmodel architecture and methodology while Section \ref{sec:experiments} discusses the results of our experimental analysis on public datasets. Finally, Section \ref{sec:conclusions} concludes the work and drafts some future work.

\section{Related Work}
\label{sec:related}

\subsection{Open Domain Q\&A}
Models most commonly used for open-domain Q\&A in IR can be broadly classified into five different families based on their architecture: Lexical models, Neural Sparse models, Late-interaction models, Re-ranking models, and Dense retrieval models. 
Lexical models include all adaptations to open-domain Q\&A of classical IR models, such as BM25~\cite{robertson1994okapi}, that do lexical matching. 
Neural Sparse models leverage deep neural networks to enhance and overcome some of the limitations of the lexical models, e.g. query-document vocabulary mismatch. They include  models such as docT5query~\cite{nogueira2019document} that uses sequence-to-sequence models to expand document terms by generating possible queries for which the document would be relevant. 
Late-interaction models rely on a bi-encoder architecture to encode the query and documents at a token level. The relevance is assessed by computing the similarity between the representations of queries terms and document terms. Late-interaction models allow the pre-computation of documents' representation by delaying the interaction between the query and document representations. 
A notable example is ColBERT~\cite{khattab2020colbert}, which computes contextualized token-level embeddings for both documents and queries and uses them at retrieval and scoring time.
Re-ranking models employ a computationally expensive neural model to re-rank documents retrieved by a fast first-stage ranker. The best-performing re-ranking model in a zero-shot retrieval scenario is currently based on a MonoT5 cross-encoder and utilizes BM25 as the first stage ranker.~\cite{rosa2022parameter}.
Dense retrieval models project the query and the documents (or passages) in a common semantic dense vector space and leverage similarity functions to score the documents according to a given query. 
Many different dense models have been recently proposed because  they empirically perform better than  lexical and sparse models in many tasks while not being computationally expensive like cross-encoder re-ranking models. 
Two dense models, namely COCO-DR~\cite{yu2022cocodr} and Contriever~\cite{izacard2021contriever}, are specifically attractive in this regard for open-domain Q\&A as they generalize very well to new domains without the need for labeled data. They are currently among the best performing dense retrieval models on the BEIR benchmarks\footnote{\href{https://docs.google.com/spreadsheets/d/1L8aACyPaXrL8iEelJLGqlMqXKPX2oSP\_R10pZoy77Ns/edit\#gid=0}{Official BEIR performance spreadsheet [Deprecated since Jan 10, 2023]}}.
Both models rely on \textit{contrastive learning}, a method that uses pairs of positive and negative examples to learn meaningful and discriminative representations for  queries and passages.
This is generally done using a synthetic dataset pseudo-labeled in a self-supervised fashion using the target domain corpus.

\subsection{Mixture-of-Experts}

In this work we employ COCO-DR and contriever in a  MoE~\cite{jacobs1991moe} framework for open-domain Q\&A. 
MoE has been used in many different contexts by the machine learning community~\cite{collobert2001moesvm,eigen2014learning,puigcerver2023sparse}.
Shazeer et al.~\cite{shazeer2017outrageously} introduced MoE in natural language processing. 
Their proposal routes a token-level representation through a fixed number of experts. 
Many works later used MoE in NLP~\cite{yann2017gatedconvolution,fedus2022switchtransformers,gaur2021mixture}.
MoE models have also been applied in the field of IR for various tasks, for example, for question answering in the biomedical domain~\cite{Dai2022MixtureOE}, visual question answering~\cite{mun2018lkdvqa}, and for rank fusion for multi-task dense retrieval~\cite{li2021-multi-task-dense}.

MoE allows the creation of expert sub-networks that specialize in an unsupervised manner and improve performance.  
Even though COCO-DR and Contriever perform exceptionally well on the BEIR benchmark, the domain knowledge is not explicitly leveraged in their training.
Due to the high domain specialization of neural networks in NLP tasks, we argue that enforcing specialized MoE IR models should yield better performance.
In this work, we rely on these pre-trained dense retrieval models and focus on improving  their performance by injecting domain specialization  based on a supervised variant of MoE.

\section{\ourmodel}
\label{sec:method}

In this section, we introduce the \ourmodel model: in Section \ref{subsec:moe_background}, we give an overview of the MoE models; in Section \ref{subsec:desire_me_model} we describe  \ourmodel, detailing its components and the training procedure, along with the differences from the classical MoE models.

\subsection{MoE background}
\label{subsec:moe_background}

Mixture-of-Experts~\cite{jacobs1991moe} (MoE) is an ensemble learning model that relies on the collective information provided by multiple expert models, which we will also call \textit{domain specializer}, or simply \textit{specializer} from hereon. Each of these specializers is dedicated to a specific topical domain or to a specific sub-task within a broader problem domain.
One of the most remarkable aspects of MoEs is their versatility as they can be employed for various types of data and tasks~\cite{collobert2001moesvm,li2021-multi-task-dense,mun2018lkdvqa}.
In the context of MoEs, a key issue is determining which specializer(s) to rely on for a specific input. 
This decision process is managed by a gating function, a significant component of a MoE model, which aims to determine the contribution of each specializer in producing the final outcome for a given input. 
The gating function is trained alongside the specializers to ensure that the gating mechanism and the specializers work together to improve the overall model's performance.
For example, let us assume to tackle a complex primary task; MoE can be employed to learn to divide it into $M$ sub-tasks, each handled by a distinct specializer. 
The gating mechanism learns to predict which sub-task  the input will likely belong to and select the appropriate specializer accordingly.

MoE operates as an ensemble model, aggregating the outputs of each specializer in a final pooling stage. 
Let $\textbf{x}$ be the vector encoding the input item and $f_i(\textbf{x})$  the output of the function, $f_i$, learned by the $i$-th specializer.
Moreover, let $g_i(\textbf{x})$ be the weight of the $i$-th  specializer  computed by the gating mechanism for input $\textbf{x}$.
Various pooling methods have been proposed in the literature to aggregate the output of the specializers.
The simplest pooling stage proposed in~\cite{zhou2022mixture}, often referred to as  \textit{Top-1 gating},
is a trivial decision model that always chooses  the output of the specializer with the highest weight, i.e.: 
\begin{align*}
    m           &= \argmax_{i=1,\hdots,M} (g_i(\textbf{x}))\\
    \textbf{y}  &= f_m(\textbf{x})
\end{align*}
Alternatively, probability scores can be derived from the gating function's output values, possibly using a \textit{softmax} normalization~\cite{jordan1994hierarchical}. The resulting probability distribution indicates the likelihood of a specializer being the most appropriate for a given input.
In this case, the pooling method makes use of the probability values from the above probability distribution as weights to compute the weighted sum of the $M$ specializers' outputs:

\begin{equation}\label{eq:weighted_moe}
    \textbf{y} = \sum_{i=1}^M f_i(\textbf{x}) \cdot g_i(\textbf{x})
\end{equation}

\subsection{The \ourmodel model}
\label{subsec:desire_me_model}

The overall structure of  \ourmodel  is very similar to that of the underlying bi-encoder dense retrieval model: we have a \textit{query encoder}, which computes the query representation, and a \textit{document encoder}, which computes the document representation. 
A scoring function, e.g., the dot product or cosine similarity, is used to compute the similarity between the dense vectors representing the query and the document. 
For efficiency purposes, the embeddings of all the documents in the collection are computed offline using the document encoder and indexed for fast retrieval.
In addition to the components of the underlying  dense retriever, we introduce in \ourmodel  a MoE module acting on the query representation only. Such a component inputs the embedding  computed by the query encoder and outputs a modified representation of the query having the same dimensionality. 
The transformation is made utilizing the \ourmodel MoE specializers detailed in the following.
Since the documents are encoded and indexed offline for fast retrieval, \ourmodel  applies the MoE only to the query representation that is typically computed online; document representations are not modified based on the specific query processed. 

\begin{figure}
    \centering    
    \begin{adjustbox}{max width=\textwidth}
        \includegraphics{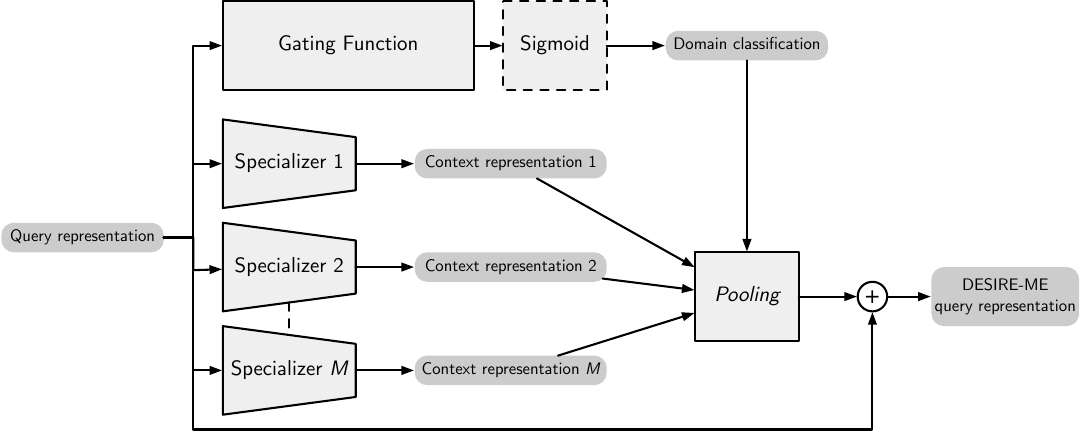}
    \end{adjustbox}
    \caption{The MoE module of the proposed model.}
    \label{fig:moe}
\end{figure}

The \ourmodel MoE  is detailed in Figure \ref{fig:moe}. The component has three major modules: the \textit{gating function}, the \textit{specializers}, and the \textit{pooling module}.

\paragraph{The gating function.}
    It  has the primary purpose  of computing the likelihood for the query to belong to any of  $M$ predefined domains. Our gating mechanism differs from  classical MoE gating functions in several ways. 
    Firstly, it relies on a multi-label domain classifier. Using a classifier as a gating function is not entirely novel in MoE; for example, in~\cite{gururangan2021demix} a Bayes posterior probability model is used to compute the output values of the gating function. Instead, we do not make the assumption of mutual exclusivity of labels, and we allow an input to belong to multiple domains. To handle multiple labels per query, we enforce that each domain is classified independently by applying a sigmoid function to the gating function output, as opposed to the commonly used softmax function. The use of softmax could compel the model to specialize even for out-of-domain queries, potentially resulting in unexpected outcomes.
    Another difference from the classical MoE models, where the gating function and the specializers' representation are trained together, is that we train end-to-end the gating function and the specializers using two distinct loss functions. While the multi-label classifier is trained using binary cross-entropy,  the MoE specializers rely on the contrastive loss computed on query-document similarity, i.e., the same loss function employed for training the underlying dense retrieval bi-encoder architecture.
    The multi-label classifier used and the process followed for generating the query labels and training it are detailed in Section \ref{sec:experiments}.

 \paragraph{The specializers.}  They  are very similar to those proposed in~\cite{jacobs1991moe}. Each of the $M$ specializers focuses on tuning the input query representation for the corresponding domain. At training time they learn via the contrastive loss function how to contextualize the query for the specific domain. 

\paragraph{The pooling module.}  
   Finally we have the \textit{pooling module} that merges the  domain context representations computed by the specializers on the basis of the domain likelihood  estimated by the gating function in the form of a normalized vector of $M$ weights. Merging is accomplished by simply  weighting and summing up the outputs of the specializers, as shown in Equation \ref{eq:weighted_moe} and depicted in Figure \ref{fig:moe}.

   We note that a consequence of the enforced domain independence condition is that an input query can be classified by our gating function as  not belonging to any of the predefined domains. 
   This is the reason why \ourmodel model has a skip connection for the input query representation that is  updated with the  domain context representation computed by the previous modules. Thanks to such a skip connection, when \ourmodel encounters an out-of-domain query, it  outputs the unmodified representation of the query not benefiting from specialization.

\section{Experimental analysis}
\label{sec:experiments}

In the following we detail the extensive experiments conducted to answer the following research questions:
\begin{enumerate}[label={\bfseries RQ\arabic*:},wide=0pt]
    \item Can  \ourmodel  enhance the effectiveness of state-of-the-art dense retrieval models for open-domain Q\&A?
    \item Does a \ourmodel model trained on a  dataset generalize to datasets having similar characteristics in a zero-shot scenario?
\end{enumerate}

\subsection{Experimental settings}
\label{subsec:experimental_settings}
In this Section, we detail the characteristics of the datasets used for the experiments; we then discuss how the datasets are used to train and test  \ourmodel. 

\paragraph{Datasets.}
In our experiments, we use four datasets included in BEIR (BEnchmarking IR~\cite{thakur2021beir}), a valuable resource for  tackling the issue of models' generalization. 
The datasets are:  NaturalQuestion~\cite{Kwiatkowski2019nq}, HotpotQA~\cite{yang2018hotpotqa}, FEVER~\cite{thorne2018fever}, and Climate-FEVER~\cite{diggelmann2021climatefever}. The main characteristics of the four datasets are resumed in Table \ref{tab:dataset_stats}. They all rely on a corpus based on Wikipedia, and provide binary relevance assessments for query-document pairs:
\begin{itemize}
 \item NaturalQuestion (NQ) contains queries submitted to the Google search engine and their answers  drawn from Wikipedia articles. 
The passages  within the Wikipedia articles that provide satisfactory answers to the questions have been identified by human annotators.

 \item HotpotQA focuses on complex questions that a single span of text might not answer and could involve reasoning over multiple documents. Queries and relevance labels have been generated with crowd-sourcing.

 \item FEVER is a resource proposed to tackle fact-checking and verification claims. It encompasses queries and documents from various domains and relies, as the previous datasets, on  a Wikipedia-based corpus.

 \item  Climate-FEVER is a dataset for verifying climate change-related claims. It includes ${\sim}1500$ test queries (no training data). The corpus is the same as FEVER, with the addition of $25$ more  documents  unavailable in  FEVER. 
\end{itemize}
\begin{table}[ht]
    \caption{Characteristics of the datasets used.  Labeled queries and the average number of labels per query refer to training queries only.}
    \label{tab:dataset_stats}    \centering
    \begin{adjustbox}{max width=\textwidth}
    \begin{tabular}{lccccccc}
    \toprule
    Dataset &
    \#Docs &
    \#Training &
    \#Validation &
    \#Test &
    Labeled docs &
    Labeled queries &
    Avg labels \\
    \midrule
    
    NaturalQuestions \cite{Kwiatkowski2019nq} &
    2,681,468 & 
    132,803 & 
    - & 
    3,452 & 
    97.1\% &
    97.8\% &
    2.04
    \\
    
    HotpotQA \cite{yang2018hotpotqa} &
    5,233,329 & 
    85,000 & 
    5,447 & 
    7,405 & 
    95.45\% &
    99.9\% &
    3.62 
    \\
    
    FEVER \cite{thorne2018fever} &
    5,416,568 & 
    109,810 & 
    6,666 & 
    6,666 & 
    91.96\% &
    99.1\% &
    2.28
    \\
    
    Climate-FEVER \cite{diggelmann2021climatefever} &
    5,416,593 & 
    - & 
    - & 
    1,535 & 
    91.95\% &
    - &
    - 
    \\
    \bottomrule
    \end{tabular}
    \end{adjustbox}
    
\end{table}

\paragraph{Query-domain labels.}
As discussed in the previous section, the \ourmodel gating function is trained in a supervised way by exploiting domain labels available for the training queries. We  automatically generated such labels for all the questions in the first three datasets by resorting to the  category assigned by contributors to their Wikipedia articles\footnote{\url{https://en.wikipedia.org/wiki/Wikipedia:FAQ/Categories}}. 
For example, the page on \textit{Eleventh Amendment to the United States Constitution} belongs to the category Law. In contrast, the page on \textit{Chinese New Year} belongs to  categories Human behavior, Culture, Society, and Religion.  
The straightforward approach we employ to create query labels involves assigning to each query the category of the corresponding Wikipedia article containing the relevant passage. However, this basic methodology proved inadequate in specific situations, necessitating the implementation of more specific actions.
The first issue arises when the relevant Wikipedia article lists specific subcategories without mentioning the main category. 
In such instances, starting from each subcategory, we navigate the Wikipedia category graph backward in a breadth-first manner until we reach the category to which the subcategory belongs. 
The second scenario occurs when the relevant article pertains to multiple categories and/or two or more Wikipedia pages are pertinent to the same query. In such cases, we identify the categories  for each page and simply label the query with all the categories of all relevant  pages.

By following this approach, we successfully label the vast majority of questions in the datasets. 
The  percentage of labeled documents and queries and the average number of per-query labels are reported in  Table~\ref{tab:dataset_stats} for the three datasets having training queries. 
The labels per query are not equally distributed: for instance, in  FEVER there are ${\sim}5000$ queries in the category \textit{Life}, meanwhile only ${\sim}500$ queries belong to the category \textit{Mathematics}. 

\paragraph{MoE specializers and training hyperparameters.}
Since in \ourmodel each specializer focuses on a specific query category, we employ $37$ distinct MoE specializers, a number equal to the number of distinct query categories in the datasets.
\ourmodel specializers feature a simple architecture: they consist of a  down-projection layer using a feed-forward network (FFN) that reduces the input dimension by half. The output layer comprises an up-projection FFN layer, which restores the vector dimension to match the input dimension.
This design draws inspiration from the adapter layer proposed in~\cite{houlsby2019parameter}. However, we opted not to use that complete adapter layer in our setup, as the skip connection is already introduced within the MoE module.
The gating function classifier has two up-projection layers, which increase the vector dimension to $2\times$ and $4\times$, respectively. The output layer is a down-projection FFN with the same size as the number of categories, i.e., $37$ in our case.
We set the training batch size to $512$, the learning rate to $10^{-5}$, and train for $60$ epochs. We use 5\% of the training set for validation and keep only the checkpoint with the lowest validation loss.

\paragraph{Metrics and baselines.} 
We assess the results of the experiments using: MAP@100, MRR@100, R@100, NDCG@10, NDCG@3 and P@1. While NDCG@10 and R@100 are commonly used on BEIR benchmarks, the additional metrics allow us to have a deeper understanding of the potential improvement of \ourmodel at small cutoffs. We also report statistically significant differences according to a Bonferroni corrected two-sided paired Student's $t$-tests with $p$-value $<0.001$. We rely on the \textit{ranx} library~\cite{bassani2022ranx} for evaluation. To  simplify comparative evaluations and to give the possibility of computing other evaluation metrics, all the runs are  made publicly available on \textit{ranxhub}\footnote{\url{https://amenra.github.io/ranxhub}}~\cite{bassani2023ranxhub}.
We compare \ourmodel variants integrated within the following different state-of-the-art dense retrieval models\footnote{Available on  HuggingFace: \href{https://huggingface.co/OpenMatch/cocodr-base-msmarco}{COCO-DR}, \href{https://huggingface.co/OpenMatch/cocodr-large-msmarco}{COCO-DR$_{XL}$} and \href{https://huggingface.co/facebook/contriever-msmarco}{Contriever}.}: COCO-DR, COCO-DR$_{XL}$, and Contriever
against the following baselines, for each dense retrieval model: 
\begin{itemize}
    \item \textit{Base.}  The original dense retrieval model without MoE in a zero-shot scenario.  
    \item \textit{Fine-tuned.} We fine-tuned the base models with the training data with a batch size of $32$ and a learning rate of $10^{-6}$ for $10$ epochs. All the other training hyper-parameters are taken from their original settings.
    \item \textit{Random\_gating (RND-G).}  We use randomly generated  weights to merge specializers' outputs. This baseline is introduced to assess the  benefits of our supervised  gating function. All other \ourmodel settings are unchanged. 
\end{itemize}

\subsection{Results and Discussion}

\paragraph{Answering RQ1.}
To answer RQ1, we conduct multiple experiments using the NQ, HotpotQA, and FEVER datasets to assess  \ourmodel capability to enhance the effectiveness of the underlying dense retrieval model. 
The results on the three datasets are reported in Table \ref{tab:nq_results}, Table \ref{tab:hotpotqa_results}, and Table \ref{tab:fever_results}, respectively.

\begin{table}[ht]
    \centering
    \caption{Results on the NQ dataset. In \textit{italic} the best results per model, in \textbf{boldface} the best results overall. Symbol * indicates a statistically significant difference over Base, Fine-tuned and RND-G.}
    \label{tab:nq_results}
    \begin{adjustbox}{max width=\textwidth}
    \begin{tabular}{lccccccc}
        \toprule
        Retriever &
        Variant &
        MAP@100 &
        MRR@100 &
        R@100 &
        NDCG@10 &
        P@1 &
        NDCG@3 \\
        \midrule  
        
        BM25 &
        - &
        0.292 &
        0.295 &
        0.758 &
        0.339 &
        0.198 &
        0.268 \\
        \midrule

        \multirow{4}*{COCO-DR} &
        
        Base &
        0.441 &
        0.455 &
        0.923 &
        0.504 &
        0.325 &
        0.424 \\

        &
        Fine-tuned &
        0.433 &
        0.446 &
        0.942 &
        0.501 &
        0.310 &
        0.411 \\

        &
        RND-G &
        0.434 &
        0.448 &
        0.926 &
        0.499 &
        0.313 &
        0.417 \\
        
        &
        \ourmodel &
        \textit{0.463}* &
        \textit{0.477}* &
        \textit{0.941} &
        \textit{0.526}* &
        \textit{0.339}* &
        \textit{0.448}* \\
        
        \midrule
        
        \multirow{4}*{Contriever} &
        
        Base &
        0.432 &
        0.446 &
        0.927 &
        0.498 &
        0.311 &
        0.414 \\
        
        & 
        Fine-tuned &
        0.427 &
        0.438 &
        0.940 &
        0.497 &
        0.295 & 
        0.406 \\
        
        &
        RND-G &
        0.441 &
        0.457 &
        0.928 &
        0.510 &
        0.320 &
        0.426 \\
        
        &
        \ourmodel &
        \textit{0.493}* &
        \textit{0.511}* &
        \textit{0.941} &
        \textit{0.559}* &
        \textit{0.379}* &
        \textit{0.480}* \\
        \midrule
        
        \multirow{4}*{COCO-DR$_{XL}$} &
        
        Base &
        0.480 &
        0.495 &
        0.937 &
        0.546 &
        0.359 &
        0.465 \\

        &
        Fine-tuned &
        0.465 &
        0.478 &
        \textit{\textbf{0.955}} &
        0.537 &
        0.331 &
        0.447 \\

        &
        RND-G &
        0.488 &
        0.503 &
        0.939 &
        0.553 &
        0.371 &
        0.473 \\
        
        &
        \ourmodel &
        \textit{\textbf{0.510}}* &
        \textit{\textbf{0.527}}* &
        0.951 &
        \textit{\textbf{0.577}}* &
        \textit{\textbf{0.390}}* &
        \textit{\textbf{0.497}}* \\
        \bottomrule
        
    \end{tabular}
    \end{adjustbox}
\end{table}


Table \ref{tab:nq_results}  reports the results of the experiments conducted with the NQ dataset.
The figures reported in the table show that fine-tuning the base model using the training data does not yield any benefit and that the integration of  \ourmodel into the  different dense retrieval systems always results in a remarkable improvement of the performances. Irrespective of the metrics considered and the dense retriever used, our solution boosts the base models of a statistically significant margin.  The Contriever relative improvement reaches an astonishing $12\%$ in NDCG@10 and 22\% in P@1 over the base model. 
This indicates that DESIRE-ME contributes significantly to enhancing the ranking quality of retrieved documents, particularly in the top positions.
Furthermore, it is also worth noting that the \algoname{RND-G} model, which relies on a random gating mechanism, does not  improve substantially the base model. This observation, which holds also for the experiments presented in the following,  proves that our gating mechanism is an important factor contributing to improved retrieval performance.

\begin{table}[ht]
    \centering
    \caption{Results on the HotpotQA dataset. In \textit{italic} the best results per model, in \textbf{boldface} the best results overall. Symbol * indicates a statistically significant difference over Base, Fine-tuned and RND-G.}
    \label{tab:hotpotqa_results}
    \begin{adjustbox}{max width=\textwidth}
    \begin{tabular}{lccccccc}
        \toprule
        Retriever &
        Variant & 
        MAP@100 &
        MRR@100 &
        R@100 &
        NDCG@10 &
        P@1 &
        NDCG@3 \\
        \midrule
        
        BM25 &
        - &
        0.521 &
        0.770 &
        0.740 &
        0.603 &
        0.707 &
        0.558 \\
        \midrule
        
        \multirow{4}*{COCO-DR}&
        
        Base &
        0.519 &
        0.795 &
        0.727 &
        0.604 &
        \textit{0.737} &
        0.563 \\

        &
        Fine-tuned &
        0.527 &
        0.753 &
        \textit{0.805} &
        0.608 &
        0.678 &
        0.553 \\
        
        &
        RND-G &
        0.523 &
        0.794 &
        0.742 &
        0.607 &
        0.734 &
        0.566 \\
        
        &
        \ourmodel &
        \textit{0.530} &
        \textit{0.795} &
        0.753 &
        \textit{0.614} &
        0.734 &
        \textit{0.571}* \\
        \midrule
        
        \multirow{4}*{Contriever}&
        Base &
        0.553 &
        0.819 &
        0.777 &
        0.638 &
        0.758 &
        0.592 \\

        &
        Fine-tuned &
        \textbf{\textit{0.575}} &
        0.799 &
        \textbf{\textit{0.848}} &
        \textbf{\textit{0.657}} &
        0.728 &
        0.600 \\
        
        &
        RND-G  &
        0.552 &
        0.817 &
        0.780 &
        0.636 &
        0.757 &
        0.592 \\
        
        &
        \ourmodel &
        0.567 &
        \textbf{\textit{0.824}} &
        0.787 &
        0.648 &
        \textbf{\textit{0.767}} &
        \textbf{\textit{0.606}} \\
        \midrule
        
        \multirow{4}*{COCO-DR$_{XL}$} &
        Base &
        0.549 &
        0.819 &
        0.756 &
        0.633 &
        0.763 &
        0.592 \\
        
        &
        Fine-tuned &
        0.542 &
        0.757 &
        \textit{0.831} &
        0.622 &
        0.681 &
        0.563 \\
        
        &
        RND-G  &
        0.555 &
        0.819 &
        0.767 &
        0.637 &
        0.763 &
        0.595 \\
        
        &
        \ourmodel &
        \textit{0.564}* &
        \textit{0.821} &
        0.780 &
        \textit{0.646}* &
        \textit{0.767} &
        \textit{0.602}* \\
        \bottomrule
        
    \end{tabular}
    \end{adjustbox}
\end{table}

In Table \ref{tab:hotpotqa_results}, we report the results on the HotpotQA dataset. 
In this case, fine-tuning the base model improves model performance, especially for R@100.
For COCO-DR and COCO-DR$_{XL}$ \ourmodel improves the performance over the baselines across all three models. The improvements are consistently statistically significant for NDCG@3. For the other metrics, except R@100, we observe a slight  improvement, but not always statistically significant. The relative performance improvement over the base model on HotpotQA is lower than that measured on NQ, reaching a margin of $3\%$ in MAP@100 and $2\%$ in NDCG@10. 
For Contriever, instead, the fine-tuned model outperforms \ourmodel in terms of R@100 and NDCG@10; meanwhile, for the other metrics \ourmodel performs slightly better than all baselines but not statistically significantly.

\begin{table}[ht]
    \centering
    \caption{Results on the FEVER dataset. In \textit{italic} the best results per model, in \textbf{boldface} the best results overall. Symbol * indicates a statistically significant difference over Base, Fine-tuned and RND-G.}
    \label{tab:fever_results}
    \begin{adjustbox}{max width=\textwidth}
    \begin{tabular}{lccccccc}
        \toprule
        Retriever &
        Variant &
        MAP@100 &
        MRR@100 &
        R@100 &
        NDCG@10 &
        P@1 &
        NDCG@3 \\
        \midrule
        
        BM25 &
        - &
        0.707 &
        0.744 &
        0.931 &
        0.753 &
        0.646 &
        0.719 \\
        \midrule
        
        \multirow{4}*{COCO-DR} &
        
        Base &
        0.660 &
        0.698 &
        0.935 &
        0.715 &
        0.586 &
        0.670 \\

        &
        Fine-tuned &
        0.544 &
        0.568 &
        0.928 &
        0.607 &
        0.431 &
        0.544 \\ 
        &
        RND-G &
        0.652 &
        0.690 &
        0.937 &
        0.710 &
        0.565 &
        0.666 \\
        
        &
        \ourmodel &
        \textit{0.696}* &
        \textit{0.736}* &
        \textit{0.945}* &
        \textit{0.749}* &
        \textit{0.623}* &
        \textit{0.712}* \\
        \midrule
        
        \multirow{4}*{Contriever}&
        
        Base &
        0.708 &
        0.749 &
        \textit{0.949} &
        0.758 &
        0.642 &
        0.724 \\

        &
        Fine-tuned &
        0.466 &
        0.483 &
        0.920 &
        0.531 &
        0.343 &
        0.458 \\
        
        &
        RND-G &
        0.709 &
        0.749 &
        0.947 &
        0.761 &
        0.640 &
        0.725 \\
        
        &
        \ourmodel &
        \textit{0.722}* &
        \textit{0.764}* &
        0.948 &
        \textit{0.772}* &
        \textit{0.655}* &
        \textit{0.739}* \\
        \midrule
        
        \multirow{4}*{COCO-DR$_{XL}$} &
        
        Base &
        0.699 &
        0.740 &
        0.946 &
        0.749 &
        0.633 &
        0.713 \\

        &
        Fine-tuned &
        0.421 & 
        0.434 & 
        0.916 &
        0.487 &
        0.296 &
        0.406 \\
        
        &
        RND-G &
        0.716 &
        0.759 &
        0.948 &
        0.765 &
        0.654 &
        0.733 \\
        
        &
        \ourmodel & 
        \textbf{\textit{0.745}}* &
        \textbf{\textit{0.789}}* &
        \textbf{\textit{0.952}} &
        \textbf{\textit{0.792}}* &
        \textbf{\textit{0.691}}* &
        \textbf{\textit{0.762}}* \\
        \bottomrule
        
    \end{tabular}
    \end{adjustbox}
\end{table}


Table \ref{tab:fever_results} shows  the performance achieved on the FEVER  dataset. 
FEVER presents a unique set of challenges compared to the other two datasets: the queries in FEVER are not questions but statements, and the relevant documents support or refute the claim made in the query statement. On this dataset, fine-tuning the base model, surprisingly, deteriorates the model performances, while BM25 performs very well, showing that the statement and the relevant documents share a similar vocabulary.
As in the previous cases, \ourmodel improves over the COCO-DR and Contriever retrievers baselines, with a relative margin of $6\%$ and $9\%$ in NDCG@10 and P@1, respectively.

It is crucial to outline that while we could replicate the COCO-DR and COCO-DR$_{XL}$ results on the NQ dataset, our results diverged slightly  from those reported in the original paper~\cite{yu2022cocodr} for  FEVER and  HotpotQA. 
The Contriever results, instead, align exactly with those reported in the original article~\cite{izacard2021contriever}. 

In summary, independently of these minor differences, our experiments on the three datasets demonstrate a consistent and significant improvement in retrieval performance obtained by integrating \ourmodel into the respective dense retrieval models. We can thus definitely answer positively RQ1.

\begin{table}[ht]
    \centering
    \caption{Results on the Climate-FEVER dataset using models trained on FEVER. In \textit{italic} the best results per model, in \textbf{boldface} the best results overall. Symbol * indicates a statistically significant difference over Base and RND-G.}
    \label{tab:climate_fever_results}
    \begin{adjustbox}{max width=\textwidth}
    \begin{tabular}{lccccccc}
        \toprule
        Retriever &
        Variant &
        MAP@100 &
        MRR@100 &
        R@100 &
        NDCG@10 &
        P@1 &
        NDCG@3 \\
        \midrule
        
        BM25 &
        - &
        0.162 &
        0.293 &
        0.436 &
        0.213 &
        0.205 &
        0.179 \\
        \midrule
        
        \multirow{4}*{COCO-DR} &
        Base &
        0.164 &
        0.290 &
        0.514 &
        0.210 &
        0.201 &
        0.171 \\
        
        &
        RND-G &
        0.170 &
        0.298 &
        0.536 &
        0.218 &
        0.207 &
        0.176 \\
        
        &
        \ourmodel &
        \textit{0.178}* &
        \textit{0.312}* &
        \textit{0.544} &
        \textit{0.228}* &
        \textit{0.219}&
        \textit{0.185}* \\
        \midrule
        
        \multirow{4}*{Contriever} &

        Base &
        0.184 &
        0.317 &
        0.574 &
        0.237 &
        0.216 &
        0.189 \\
        
        &
        RND-G &
        \textbf{\textit{0.205}} &
        0.351 &
        \textbf{\textit{0.609}} &
        0.264 &
        0.241 &
        0.211 \\
        
        &
        \ourmodel &
        \textbf{\textit{0.205}} &
        \textbf{\textit{0.358}} &
        0.600 &
        \textbf{\textit{0.268}} &
        \textbf{\textit{0.250}} &
        \textbf{\textit{0.213}} \\
        \midrule
        
        \multirow{4}*{COCO-DR$_{XL}$} &
      
        Base &
        0.180 &
        0.322 &
        0.547 &
        0.231 &
        0.227 &
        0.189 \\
        
        &
        RND-G &
        0.182 &
        0.325 &
        0.564 &
        0.234 &
        0.229 &
        0.188 \\
        
        &
        \ourmodel &
        \textit{0.191}* &
        \textit{0.343}* &
        \textit{0.573} &
        \textit{0.247}* &
        \textit{0.243} &
        \textit{0.199}* \\
        \bottomrule
        
    \end{tabular}
    \end{adjustbox}
\end{table}

\paragraph{Answering RQ2.}
We  evaluate \ourmodel trained on FEVER in a zero-shot scenario on Climate-FEVER. This experiments aims to assess the generalization power of \ourmodel  on a similar yet distinct dataset.
Climate-FEVER and FEVER share a substantial portion of their corpus. 
However, an important distinction lies in the queries: Climate-FEVER relies on real-world user queries, while FEVER employs synthetic queries.
We report in Table \ref{tab:climate_fever_results} the results of the experiments conducted using the \ourmodel models trained on the FEVER  on the questions of Climate-FEVER. 
Despite the difference in query types, we  notice improvements over the baselines across all models, similar to the previous three experiments. Specifically, the improvements over the respective base models are statistically significant for all the metrics measured with both COCO-DR retrievers. The relative margin in terms of NDCG@10 reaches 9\%.
These results outlines the capacity of \ourmodel to adapt to  incoming queries that differs substantially from the ones seen at training time. We can thus answer positively also the second research question (RQ2) even if further experiments involving also other corpora are needed to undoubtedly assess  the generalization power of \ourmodel across totally different Q\&A scenarios.

\section{Conclusions}
\label{sec:conclusions}
In this work we introduced \ourmodel, a new retrieval model for open-domain Q\&A task that leverages the Mixture-of-Experts (MoE) framework to improve the performance of state-of-the-art dense retrieval models. 
The proposed MoE component uses supervised methods in the gating mechanism and predicts the likelihood of a query belonging to predefined domains, while the specializer modules focus on contextualizing the query vector for specific domains.
We conducted extensive experiments across multiple datasets to investigate two research questions.
For the first experiment, we chose three diverse datasets.
Our experiments show that integrating the \ourmodel model into dense retrieval models leads to significant improvements in various retrieval metrics, answering positively the \textbf{RQ1}. 
These findings highlight the robustness and adaptability of \ourmodel.
In response to the \textbf{RQ2}, the experiment performed on the Climate-FEVER dataset, using a model trained on FEVER shows that MoE can generalize to new datasets in a zero-shot scenario. 
This also shows the potential of leveraging knowledge from a similar corpus and encourages further exploration of techniques, such as transfer learning in the open-domain Q\&A tasks.

\noindent
\paragraph{Limitations and future work.} 
Our primary focus was understanding the improvements achieved by using domain specialization in open-domain Q\&A; we did not concentrate on optimizing the underlying neural architectures for the specializers and gating mechanism.
The main limitation of this work is the assumption of having query domain information, which might not be true in most IR tasks. 
In our experiments, we relied on Wikipedia corpora and categories; our labeling process is however not exportable to other cases. 
Consequently, given the diversity in real-world queries and documents our insights could be not directly generalizable to other settings.
Future research could address this issue by evaluating \ourmodel on more diverse and extensive datasets.
This would require extensive user studies and crowd-sourcing to label query domains or topics.
Another option would be using LLMs to create soft labels for queries~\cite{hashemi2023dense}.
Another future research topic is query augmentation, which can be addressed by adapting the \ourmodel specializer modules to domain-specific query expansion modules. 
This way, the query expansion would occur by using models that can leverage domain-specific vocabularies.   

\section*{Acknowledgements}
Funding for this research has been provided by  spoke ``FutureHPC \& BigData'' of the ICSC – Centro Nazionale di Ricerca in High-Performance Computing, Big Data and Quantum Computing,  Spoke ``Human-centered AI'' of the M4C2 - Investimento 1.3, Partenariato Esteso PE00000013 - "FAIR - Future Artificial Intelligence Research", the FoReLab project (Departments of Excellence),  the NEREO and CAMEO PRIN projects funded by the Italian Ministry of Education and Research Grant no. 2022AEFHAZ and 2022ZLL7MW, and the EFRA project funded by the European Commission under the NextGeneration EU programme grant agreement n. 101093026. However, the views and opinions expressed are those of the authors only and do not necessarily reflect those of the EU or European Commission-EU. Neither the EU nor the granting authority can be held responsible for them.

\bibliographystyle{splncs04}
\bibliography{biblio}

\begin{thebibliography}{10}
\providecommand{\url}[1]{\texttt{#1}}
\providecommand{\urlprefix}{URL }
\providecommand{\doi}[1]{https://doi.org/#1}

\bibitem{bassani2022ranx}
Bassani, E.: ranx: {A} blazing-fast python library for ranking evaluation and comparison. In: Hagen, M., Verberne, S., Macdonald, C., Seifert, C., Balog, K., N{\o}rv{\aa}g, K., Setty, V. (eds.) Advances in Information Retrieval - 44th European Conference on {IR} Research, {ECIR} 2022, Stavanger, Norway, April 10-14, 2022, Proceedings, Part {II}. Lecture Notes in Computer Science, vol. 13186, pp. 259--264. Springer (2022). \doi{10.1007/978-3-030-99739-7\_30}, \url{https://doi.org/10.1007/978-3-030-99739-7\_30}

\bibitem{bassani2023ranxhub}
Bassani, E.: Ranxhub: An online repository for information retrieval runs. In: Proceedings of the 46th International ACM SIGIR Conference on Research and Development in Information Retrieval. p. 3210–3214. SIGIR '23, Association for Computing Machinery, New York, NY, USA (2023). \doi{10.1145/3539618.3591823}, \url{https://doi.org/10.1145/3539618.3591823}

\bibitem{collobert2001moesvm}
Collobert, R., Bengio, S., Bengio, Y.: A parallel mixture of svms for very large scale problems. In: Dietterich, T., Becker, S., Ghahramani, Z. (eds.) Advances in Neural Information Processing Systems. vol.~14. MIT Press (2001), \url{https://proceedings.neurips.cc/paper\_files/paper/2001/file/36ac8e558ac7690b6f44e2cb5ef93322-Paper.pdf}

\bibitem{Dai2022MixtureOE}
Dai, D., Jiang, W.J., Zhang, J., Peng, W., Lyu, Y., Sui, Z., Chang, B., Zhu, Y.: Mixture of experts for biomedical question answering. ArXiv  \textbf{abs/2204.07469} (2022), \url{https://api.semanticscholar.org/CorpusID:248218762}

\bibitem{yann2017gatedconvolution}
Dauphin, Y.N., Fan, A., Auli, M., Grangier, D.: Language modeling with gated convolutional networks. In: Precup, D., Teh, Y.W. (eds.) Proceedings of the 34th International Conference on Machine Learning. Proceedings of Machine Learning Research, vol.~70, pp. 933--941. PMLR (06--11 Aug 2017), \url{https://proceedings.mlr.press/v70/dauphin17a.html}

\bibitem{diggelmann2021climatefever}
Diggelmann, T., Boyd-Graber, J., Bulian, J., Ciaramita, M., Leippold, M.: Climate-fever: A dataset for verification of real-world climate claims (2021)

\bibitem{eigen2014learning}
Eigen, D., Ranzato, M., Sutskever, I.: Learning factored representations in a deep mixture of experts (2014)

\bibitem{fedus2022switchtransformers}
Fedus, W., Zoph, B., Shazeer, N.: Switch transformers: Scaling to trillion parameter models with simple and efficient sparsity. J. Mach. Learn. Res.  \textbf{23}(1) (jan 2022)

\bibitem{gaur2021mixture}
Gaur, N., Farris, B., Haghani, P., Leal, I., Moreno, P.J., Prasad, M., Ramabhadran, B., Zhu, Y.: Mixture of informed experts for multilingual speech recognition. In: ICASSP 2021 - 2021 IEEE International Conference on Acoustics, Speech and Signal Processing (ICASSP). pp. 6234--6238 (June 2021). \doi{10.1109/ICASSP39728.2021.9414379}

\bibitem{gururangan2021demix}
Gururangan, S., Lewis, M., Holtzman, A., Smith, N.A., Zettlemoyer, L.: Demix layers: Disentangling domains for modular language modeling (2021)

\bibitem{hashemi2023dense}
Hashemi, H., Zhuang, Y., Kothur, S.S.R., Prasad, S., Meij, E., Croft, W.B.: Dense retrieval adaptation using target domain description (2023)

\bibitem{houlsby2019parameter}
Houlsby, N., Giurgiu, A., Jastrzebski, S., Morrone, B., De~Laroussilhe, Q., Gesmundo, A., Attariyan, M., Gelly, S.: Parameter-efficient transfer learning for {NLP}. In: Proceedings of the 36th International Conference on Machine Learning (2019)

\bibitem{izacard2021contriever}
Izacard, G., Caron, M., Hosseini, L., Riedel, S., Bojanowski, P., Joulin, A., Grave, E.: Unsupervised dense information retrieval with contrastive learning (2021). \doi{10.48550/ARXIV.2112.09118}, \url{https://arxiv.org/abs/2112.09118}

\bibitem{jacobs1991moe}
Jacobs, R.A., Jordan, M.I., Nowlan, S.J., Hinton, G.E.: Adaptive mixtures of local experts. Neural Computation  \textbf{3}(1),  79--87 (March 1991). \doi{10.1162/neco.1991.3.1.79}

\bibitem{jordan1994hierarchical}
Jordan, M.I., Jacobs, R.A.: Hierarchical mixtures of experts and the em algorithm. Neural Computation  \textbf{6}(2),  181--214 (1994). \doi{10.1162/neco.1994.6.2.181}

\bibitem{khattab2020colbert}
Khattab, O., Zaharia, M.: Colbert: Efficient and effective passage search via contextualized late interaction over bert. In: Proceedings of the 43rd International ACM SIGIR Conference on Research and Development in Information Retrieval. p. 39–48. SIGIR '20, Association for Computing Machinery, New York, NY, USA (2020). \doi{10.1145/3397271.3401075}, \url{https://doi.org/10.1145/3397271.3401075}

\bibitem{Kwiatkowski2019nq}
Kwiatkowski, T., Palomaki, J., Redfield, O., Collins, M., Parikh, A., Alberti, C., Epstein, D., Polosukhin, I., Kelcey, M., Devlin, J., Lee, K., Toutanova, K.N., Jones, L., Chang, M.W., Dai, A., Uszkoreit, J., Le, Q., Petrov, S.: Natural questions: a benchmark for question answering research. Transactions of the Association of Computational Linguistics  (2019)

\bibitem{li2021-multi-task-dense}
Li, M., Li, M., Xiong, K., Lin, J.: Multi-task dense retrieval via model uncertainty fusion for open-domain question answering. In: Findings of the Association for Computational Linguistics: EMNLP 2021. pp. 274--287. Association for Computational Linguistics, Punta Cana, Dominican Republic (Nov 2021). \doi{10.18653/v1/2021.findings-emnlp.26}, \url{https://aclanthology.org/2021.findings-emnlp.26}

\bibitem{mitra2019nir}
Mitra, B., Craswell, N.: An introduction to neural information retrieval. Foundations and Trends® in Information Retrieval  \textbf{13}(1),  1--126 (2018). \doi{10.1561/1500000061}, \url{http://dx.doi.org/10.1561/1500000061}

\bibitem{mun2018lkdvqa}
Mun, J., Lee, K., Shin, J., Han, B.: Learning to specialize with knowledge distillation for visual question answering. In: Bengio, S., Wallach, H., Larochelle, H., Grauman, K., Cesa-Bianchi, N., Garnett, R. (eds.) Advances in Neural Information Processing Systems. vol.~31. Curran Associates, Inc. (2018), \url{https://proceedings.neurips.cc/paper\_files/paper/2018/file/0f2818101a7ac4b96ceeba38de4b934c-Paper.pdf}

\bibitem{nogueira2019document}
Nogueira, R., Yang, W., Lin, J., Cho, K.: Document expansion by query prediction (2019)

\bibitem{puigcerver2023sparse}
Puigcerver, J., Riquelme, C., Mustafa, B., Houlsby, N.: From sparse to soft mixtures of experts (2023)

\bibitem{robertson1994okapi}
Robertson, S.E., Walker, S., Jones, S., Hancock{-}Beaulieu, M., Gatford, M.: Okapi at {TREC-3}. In: Harman, D.K. (ed.) Proceedings of The Third Text REtrieval Conference, {TREC} 1994, Gaithersburg, Maryland, USA, November 2-4, 1994. {NIST} Special Publication, vol. 500-225, pp. 109--126. National Institute of Standards and Technology {(NIST)} (1994), \url{http://trec.nist.gov/pubs/trec3/papers/city.ps.gz}

\bibitem{rosa2022parameter}
Rosa, G.M., Bonifacio, L., Jeronymo, V., Abonizio, H., Fadaee, M., Lotufo, R., Nogueira, R.: No parameter left behind: How distillation and model size affect zero-shot retrieval (2022)

\bibitem{shazeer2017outrageously}
Shazeer, N., Mirhoseini, A., Maziarz, K., Davis, A., Le, Q., Hinton, G., Dean, J.: Outrageously large neural networks: The sparsely-gated mixture-of-experts layer (2017)

\bibitem{thakur2021beir}
Thakur, N., Reimers, N., R{\"u}ckl{\'e}, A., Srivastava, A., Gurevych, I.: {BEIR}: A heterogeneous benchmark for zero-shot evaluation of information retrieval models. In: Thirty-fifth Conference on Neural Information Processing Systems Datasets and Benchmarks Track (Round 2) (2021), \url{https://openreview.net/forum?id=wCu6T5xFjeJ}

\bibitem{thorne2018fever}
Thorne, J., Vlachos, A., Christodoulopoulos, C., Mittal, A.: {FEVER}: a large-scale dataset for fact extraction and {VER}ification. In: Proceedings of the 2018 Conference of the North {A}merican Chapter of the Association for Computational Linguistics: Human Language Technologies, Volume 1 (Long Papers). pp. 809--819. Association for Computational Linguistics, New Orleans, Louisiana (Jun 2018). \doi{10.18653/v1/N18-1074}, \url{https://aclanthology.org/N18-1074}

\bibitem{yang2018hotpotqa}
Yang, Z., Qi, P., Zhang, S., Bengio, Y., Cohen, W., Salakhutdinov, R., Manning, C.D.: {H}otpot{QA}: A dataset for diverse, explainable multi-hop question answering. In: Proceedings of the 2018 Conference on Empirical Methods in Natural Language Processing. pp. 2369--2380. Association for Computational Linguistics, Brussels, Belgium (Oct-Nov 2018). \doi{10.18653/v1/D18-1259}, \url{https://aclanthology.org/D18-1259}

\bibitem{yu2022cocodr}
Yu, Y., Xiong, C., Sun, S., Zhang, C., Overwijk, A.: Coco-dr: Combating distribution shifts in zero-shot dense retrieval with contrastive and distributionally robust learning. In: Proceedings of the 2022 Conference on Empirical Methods in Natural Language Processing. pp. 1462--1479 (2022)

\bibitem{zhou2022mixture}
Zhou, Y., Lei, T., Liu, H., Du, N., Huang, Y., Zhao, V., Dai, A.M., Le, Q.V., Laudon, J., et~al.: Mixture-of-experts with expert choice routing. Advances in Neural Information Processing Systems  \textbf{35},  7103--7114 (2022)

\end{thebibliography}

\end{document}